\documentclass[prl,reprint,superscriptaddress,preprintnumbers,nofootinbib,sort&compress]{revtex4-1}
\usepackage{graphicx}
\usepackage[T1]{fontenc}
\usepackage{color}
\usepackage{amsmath}
\usepackage{amssymb}
\usepackage{dcolumn}
\graphicspath{{../Figures/}{.}}

\newcommand{\unige}{\affiliation{Universit\`a degli Studi di Genova, Via Dodecaneso 33, 16146 Genova, Italy}}
\newcommand{\infnge}{\affiliation{Istituto Nazionale di Fisica Nucleare (INFN), Sezione di Genova, Via Dodecaneso 33, 16146 Genova, Italy}}
\newcommand{\hzdr}{\affiliation{Helmholtz-Zentrum Dresden-Rossendorf, Bautzner Landstr. 400, 01328 Dresden, Germany}}
\newcommand{\tudd}{\affiliation{Technische Universit\"at Dresden, Institut f\"ur Kern- und Teilchenphysik, Zellescher Weg 19, 01069 Dresden, Germany}}
\newcommand{\unipd}{\affiliation{Dipartimento di Fisica e Astronomia, Universit\`a degli Studi di Padova, Via F. Marzolo 8, 35131 Padova, Italy}}
\newcommand{\infnpd}{\affiliation{INFN, Sezione di Padova, Via F. Marzolo 8, 35131 Padova, Italy}}
\newcommand{\edin}{\affiliation{SUPA, School of Physics and Astronomy, University of Edinburgh, EH9 3FD Edinburgh, United Kingdom}}
\newcommand{\unina}{\affiliation{Universit\`a degli Studi di Napoli "Federico II", Dipartimento di Fisica "E. Pancini", Via Cintia, 80126 Napoli, Italy}}
\newcommand{\infnna}{\affiliation{INFN, Sezione di Napoli, Via Cintia, 80126 Napoli, Italy}}
\newcommand{\gssi}{\affiliation{Gran Sasso Science Institute, 67100 L'Aquila, Italy}}
\newcommand{\lngs}{\affiliation{INFN Laboratori Nazionali del Gran Sasso (LNGS), 67100 Assergi (AQ), Italy}}
\newcommand{\uniba}{\affiliation{Universit\`a degli Studi di Bari, 70125 Bari, Italy}}
\newcommand{\infnba}{\affiliation{INFN, Sezione di Bari, 70125 Bari, Italy}}
\newcommand{\atomki}{\affiliation{Institute for Nuclear Research (MTA ATOMKI), PO Box 51, H-4001 Debrecen, Hungary}}
\newcommand{\unito}{\affiliation{Universit\`a degli Studi di Torino, Via P. Giuria 1, 10125 Torino, Italy}}
\newcommand{\infnto}{\affiliation{INFN, Sezione di Torino, Via P. Giuria 1, 10125 Torino, Italy}}
\newcommand{\unimi}{\affiliation{Universit\`a degli Studi di Milano, Via G. Celoria 16, 20133 Milano, Italy}}
\newcommand{\infnmi}{\affiliation{INFN, Sezione di Milano, Via G. Celoria 16, 20133 Milano, Italy}}
\newcommand{\infnrm}{\affiliation{INFN, Sezione di Roma La Sapienza, Piazzale A. Moro 2, 00185 Roma, Italy}}
\newcommand{\buda}{\affiliation{Konkoly Observatory, Research Centre for Astronomy and Earth Sciences, Hungarian Academy of Sciences, 1121 Budapest, Hungary}}
\newcommand{\teramo}{\affiliation{Osservatorio Astronomico di Collurania, Teramo, Italy}}

\newcommand{\monash}{\affiliation{Monash Centre for Astrophysics, School of Physics \& Astronomy, Monash University, VIC 3800, Australia}}

\begin{document}
\title{Direct capture cross section and the $E_p$ = 71 and 105 keV resonances in the $^{22}$Ne($p,\gamma$)$^{23}$Na reaction}

\author{F.\,Ferraro}\unige\infnge
\author{M.\,P.\,Tak\'acs}\hzdr\tudd
\author{D.\,Piatti}\unipd\infnpd
\author{F.\,Cavanna}\infnge
\author{R.\,Depalo}\unipd\infnpd
\author{M.\,Aliotta}\edin
\author{D.\,Bemmerer}\email[e-mail address:\,]{d.bemmerer@hzdr.de}\hzdr
\author{A.\,Best}\unina\infnna
\author{A.\,Boeltzig}\gssi
\author{C.\,Broggini}\infnpd
\author{C.G.\,Bruno}\edin
\author{A.\,Caciolli}\email[e-mail address:\,]{caciolli@pd.infn.it}\unipd\infnpd
\author{T.\,Chillery}\edin
\author{G.\,F.\,Ciani}\gssi\lngs
\author{P.\,Corvisiero}\unige\infnge
\author{T.\,Davinson}\edin
\author{G.\,D'Erasmo}\uniba\infnba
\author{A.\,Di\,Leva}\unina\infnna
\author{Z.\,Elekes}\atomki
\author{E.\,M.\,Fiore}\uniba\infnba
\author{A.\,Formicola}\lngs
\author{Zs.\,F\"ul\"op}\atomki
\author{G.\,Gervino}\infnto\unito
\author{A.\,Guglielmetti}\unimi\infnmi
\author{C.\,Gustavino}\infnrm
\author{Gy.\,Gy\"urky}\atomki
\author{G.\,Imbriani}\unina\infnna
\author{M.\,Junker}\lngs
\author{A.\,Karakas}\monash
\author{I.\,Kochanek}\lngs
\author{M.\,Lugaro}\buda
\author{P.\,Marigo}\unipd\infnpd
\author{R.\,Menegazzo}\infnpd
\author{V.\,Mossa}\uniba\infnba
\author{F. R.\,Pantaleo}\uniba\infnba
\author{V.\,Paticchio}\infnba
\author{R.\,Perrino} \altaffiliation[Permanent address: ]{INFN Sezione di Lecce, Lecce, Italy} \infnba
\author{P.\,Prati}\unige\infnge
\author{L.\,Schiavulli}\uniba\infnba
\author{K.\,St\"ockel}\hzdr\tudd
\author{O.\,Straniero}\teramo\infnna
\author{T.\,Sz\"ucs}\hzdr\atomki
\author{D.\,Trezzi}\unimi\infnmi
\author{S.\,Zavatarelli}\infnge
\collaboration{The LUNA Collaboration}\noaffiliation

\begin{abstract}

The $^{22}$Ne($p,\gamma$)$^{23}$Na reaction, part of the neon-sodium cycle of hydrogen burning, may explain the observed anticorrelation between sodium and oxygen abundances in globular cluster stars. Its rate is controlled by a number of low-energy resonances and a slowly varying non-resonant component. Three new resonances at $E_p$ = 156.2, 189.5, and 259.7 keV have recently been observed and confirmed. However, significant uncertainty on the reaction rate remains due to the non-resonant process and to two suggested resonances at $E_p$ = 71 and 105 keV. Here, new $^{22}$Ne($p,\gamma$)$^{23}$Na data with high statistics and low background are reported. Stringent upper limits of 6$\times$10$^{-11}$ and 7$\times$10$^{-11}$\,eV (90\% confidence level), respectively, are placed on the two suggested resonances. In addition, the off-resonant S-factor has been measured at unprecedented low energy, constraining the contributions from a subthreshold resonance and the direct capture process. As a result, at a temperature of 0.1 GK the error bar of the $^{22}$Ne($p,\gamma$)$^{23}$Na rate is now reduced by three orders of magnitude. 

\end{abstract}


\maketitle

{\bf Introduction.} At the base of the convective envelope of massive (initial stellar mass $M$ $\gtrsim4M_\odot$, where $M_\odot$ is the mass of the Sun) asymptotic giant branch (AGB) stars, temperatures as high as $T_9\sim$ 0.1 ($T_9$ is the stellar temperature in 10$^9$\,K) can occur, facilitating the so-called hot bottom burning (HBB) process. Also in massive stars of $M\gtrsim 50 M_\odot$, the ashes of hydrogen burning at temperatures up to $T_9\sim$ 0.08 can be exposed to the stellar surface due to the strong winds, see e.g. \cite{Decressin07-AA}. As a result, in addition to the carbon-nitrogen-oxygen (CNO) cycle of hydrogen burning \cite{Broggini18-PPNP}, also more advanced processes are operating \cite{Slemer17-MNRAS}:  the neon-sodium (NeNa) and magnesium-aluminum (MgAl) cycles \cite{Denissenkov15-MNRAS}. 

Within the NeNa cycle, the $^{22}$Ne($p,\gamma$)$^{23}$Na reaction links $^{22}$Ne, the third most abundant nuclide produced in stellar helium burning \cite{Buchmann06-NPA} and an important neutron source for the astrophysical s-process \cite{Kaeppeler11-RMP}, to $^{23}$Na, the only stable isotope of sodium. This reaction is responsible for the  anticorrelation of oxygen and sodium abundances observed in  globular clusters \cite{Carretta09-AA,Gratton12-AAR} and  its rate affects models seeking to reproduce this anticorrelation, see Ref. \cite{Ventura18-MNRAS} for a recent example.

Until recently, the $^{22}$Ne($p,\gamma$)$^{23}$Na rate was very uncertain, with a discrepancy of a factor of 1000 between the recommended rates from the NACRE \cite{NACRE99-NPA} compilation, on the one hand, and the evaluations by Hale {\it et al.} \cite{Hale01-PRC}, Iliadis {\it et al.} \cite{Iliadis10-NPA841_31}, and STARLIB \cite{Sallaska13-ApJSS}, on the other hand. This situation dramatically changed with the first observation of three low-energy resonances in an experiment using high-purity germanium (HPGe) detectors \cite{Cavanna15-PRL,Depalo16-PRC,Cavanna15-PRL-Erratum,Bemmerer18-EPL}. The data were taken at the Laboratory for Underground Nuclear Astrophysics (LUNA) \cite{Broggini18-PPNP} as part of a broader campaign of underground, direct cross section measurements \cite{Bemmerer06-PRL,Scott12-PRL,Anders14-PRL,Bruno16-PRL,Lugaro17-NA}. Remarkably, two of the three new $^{22}$Ne($p,\gamma$)$^{23}$Na resonance strengths \cite{Cavanna15-PRL,Depalo16-PRC,Cavanna15-PRL-Erratum,Bemmerer18-EPL} are higher than the values or the upper limits  previously derived \cite{NACRE99-NPA,Iliadis10-NPA841_251}  from indirect methods \cite{Hale01-PRC}, underlining the importance of direct data.

Recently, the existence of the two lowest out of the three new resonances at $E_p$ = 156.2, 189.5, and 259.7 keV ($E_p$ denotes the proton beam energy in the laboratory system, $E$ the center of mass energy) was independently confirmed in a surface-based experiment, even though that work reported higher experimental background and 5-11\%  higher absolute resonance strengths \cite{Kelly17-PRC}.  Similarly, preliminary data at the DRAGON recoil separator confirmed the existence of the two highest out of the three new resonances \cite{Lennarz17-NPA8}.  

However, in order to precisely understand the thermonuclear reaction rate at very low temperatures, $T_9$ $\leq$ 0.1, two pieces of the $^{22}$Ne($p,\gamma$)$^{23}$Na puzzle are still missing.  The present work aims to address both of them: first, two resonances at even lower energy, $E_p$ = 71 and 105\,keV had been reported as tentative in an early indirect work \cite{Powers71-PRC} but were not confirmed in later studies \cite{Hale01-PRC,Jenkins13-PRC}. The HPGe-based LUNA experiment provided new upper limits \cite{Cavanna15-PRL,Depalo16-PRC}, but could not fully exclude them. Second, the contributions from direct capture and broad resonances were not addressed in recent work at LUNA \cite{Cavanna15-PRL,Depalo16-PRC,Cavanna15-PRL-Erratum,Bemmerer18-EPL} or elsewhere \cite{Kelly17-PRC,Lennarz17-NPA8}.  

%==============================================
\begin{figure}[tb]
%\centering
\includegraphics[width=\columnwidth]{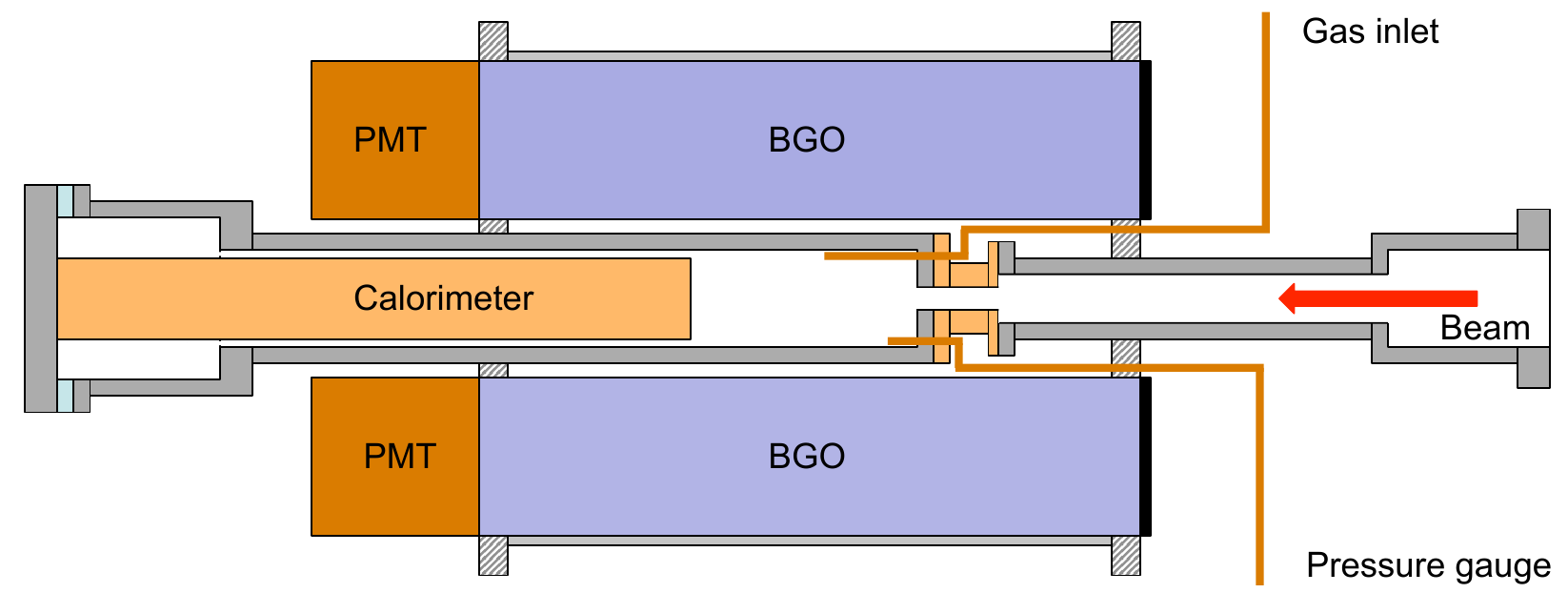}
\caption{Sketch of the experimental setup \cite{Ferraro18-EPJA}. The ion beam entered from the right after passing through three consecutive differential pumping stages (not shown) and was stopped on the massive copper beam calorimeter with constant temperature gradient.}
\label{fig:Setup}
\end{figure}
%==============================================

{\bf Setup.} For the irradiations, a windowless gas target chamber filled with 99.9\% isotopically enriched $^{22}$Ne gas was used \cite{Ferraro18-EPJA,Ferraro17-PhD,Takacs17-PhD}. The target chamber was a stainless steel tube of 56\,mm inner diameter, with a central part of 108\,mm length that was filled with 2.0\,mbar $^{22}$Ne gas and placed at the center of a 280\,mm long bismuth germanate (BGO) borehole detector of 60 (200) mm inner (outer) diameter (Figure\,\ref{fig:Setup}). 

The $E_p$ = 66-310\,keV, 60-260\,$\mu$A H$^+$ beam from the LUNA 400\,kV accelerator \cite{Formicola03-NIMA} was bent by 45$^\circ$ and then passed three consecutive pumping stages with typical pressures of 5$\times$10$^{-7}$, 2$\times$10$^{-6}$, and 5$\times$10$^{-3}$\,mbar. The exhaust from the pumps was collected, purified in a chemical getter, and recirculated \cite{Ferraro18-EPJA}. The main contaminant of the gas, nitrogen, was monitored by runs on top of the $E_p$ = 278\,keV $^{14}$N(p,$\gamma$)$^{15}$O resonance and kept below 1\% (by volume). The beam intensity was measured to 1.5\% precision with a beam calorimeter \cite{Ferraro18-EPJA}.

The scintillation light from each of the six longitudinally separated optical segments of the BGO detector was detected by a photomultiplier, preamplified, passed to one channel of a CAEN V1724 100 Ms/s digitizer module, with a separate digital trigger for each channel, timestamped and stored in list mode for offline analysis. 
After gain matching using the laboratory radioactive background and correcting for gain drifts of up to 2\%/day, two type of spectra were reconstructed offline: the sum of the six single spectra and the add-back spectrum that is equivalent to using the whole BGO as a single detector \cite{Ferraro18-EPJA}.
A pulser was used to determine the dead time, 
3\% or less in the runs with $^{22}$Ne gas. 

\begin{figure*}[tb]
\includegraphics[width=\textwidth]{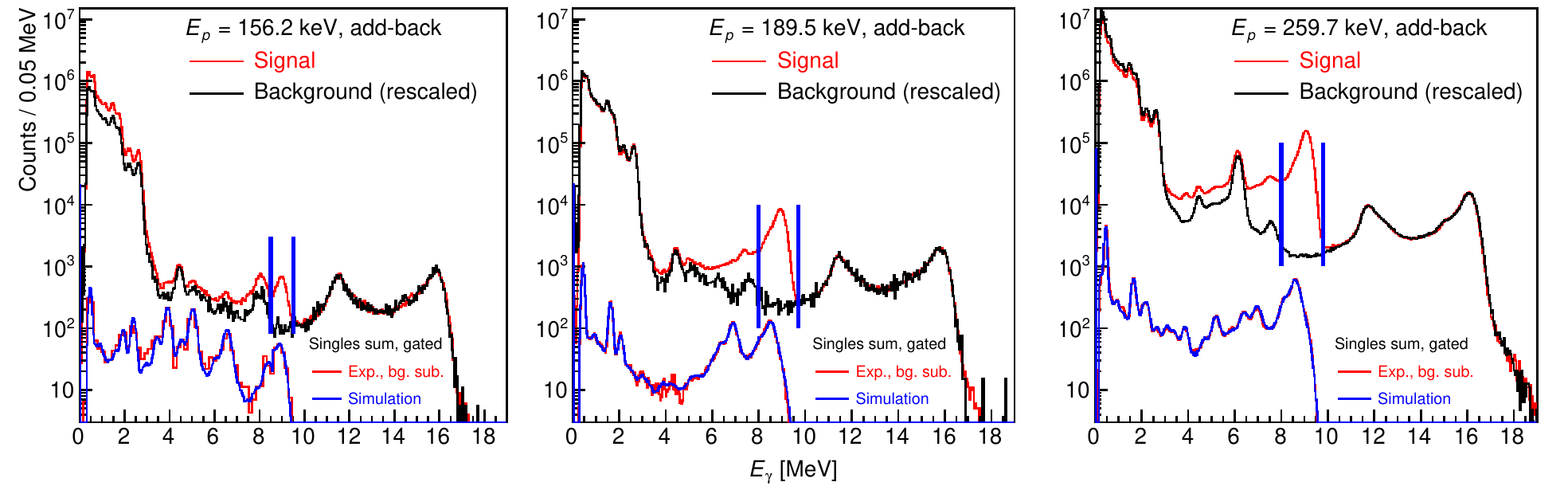}%
\caption{\label{fig:Gammaspectra} In-beam $\gamma$-ray spectra on top of the three resonances at $E_p$ = 156.2, 189.5, and 259.7 keV, and the rescaled background (see text). The ROI for the resonance strength determination is marked with vertical bands. The singles-sum spectra, background subtracted (bg. subtr.) and gated on add-back counts in the ROI, are also shown, together with the simulation.}
\end{figure*}

{\bf Experiment.} %
For each of the three previously discovered resonances \cite{Cavanna15-PRL,Depalo16-PRC,Cavanna15-PRL-Erratum,Bemmerer18-EPL}, first the proton beam energy $E_p$ was varied in 1-3 keV steps, performing short runs at each $E_p$ value. Second, a long run was performed at the energy with the highest yield, and the spectrum of emitted $\gamma$ rays was analyzed (Figure~\ref{fig:Gammaspectra}). 
The Compton continuum from the $^{11}$B($p,\gamma$)$^{12}$C background reaction was subtracted based on a run with inert argon gas at a pressure that gave the same proton beam energy loss as neon gas at this energy \cite{Ferraro18-EPJA}. The argon run was scaled for equal counting rate in the 10-19\,MeV region, where the $^{11}$B($p,\gamma$)$^{12}$C high-energy $\gamma$ rays dominate and the shapes of the two spectra agree (Figure~\ref{fig:Gammaspectra}). In order to check for unsubtracted Compton background in the region of interest (ROI), the singles spectra, gated on add-back counts in the ROI and background subtracted as described above, were compared with GEANT4 \cite{Agostinelli03-NIMA} and GEANT3 \cite{Arpesella95-NIMA} Monte Carlo simulations using the previously reported branching ratios \cite{Depalo16-PRC,Ferraro18-EPJA}. Experimental and simulated singles spectra agree (Figure~\ref{fig:Gammaspectra}).

\begin{figure*}[tb]
%\centering
\includegraphics[width=\textwidth]{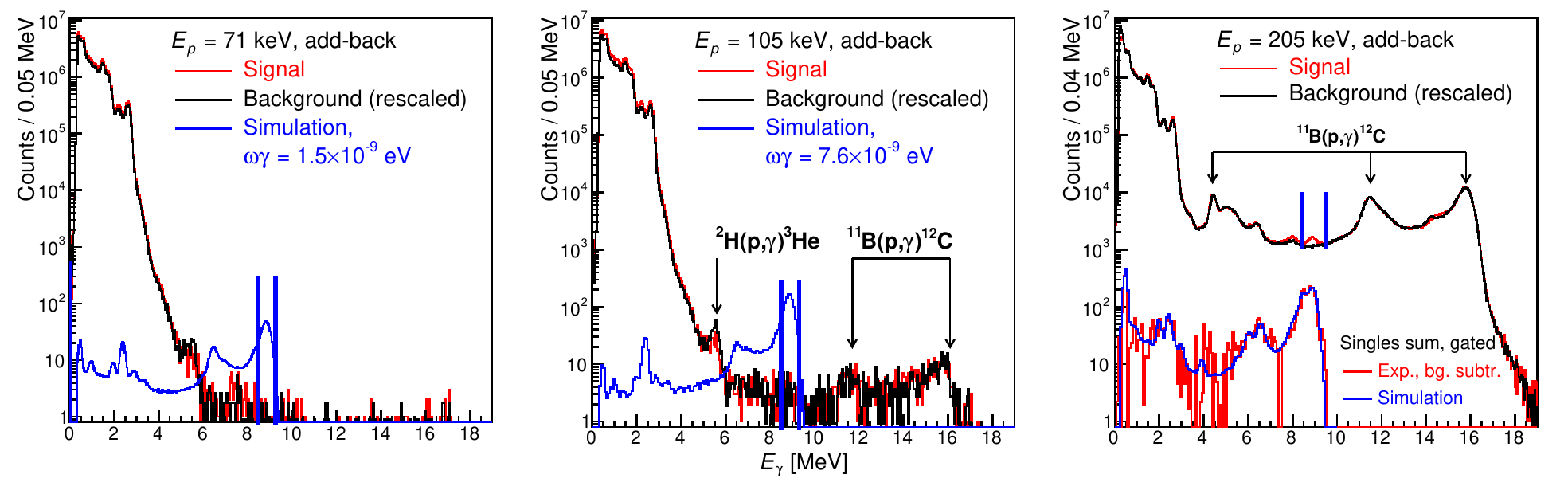}
\caption{\label{fig:Gammaspec76} Experimental add-back $\gamma$-ray spectra (red line), experimental background (black line, rescaled), and simulation (blue line). Runs on top of the suggested resonances at $E_p$ = 71 and 105\,keV (left and middle panel, respectively), and for direct capture (right panel). In the right panel, the gated and background subtracted singles-sum spectrum is also included.}
\end{figure*}

Finally, for the two suggested resonances at $E_p$ = 71 and 105\,keV, several long runs that covered a range from 63-78 and 95-113\,keV, respectively, were performed (Figure~\ref{fig:Gammaspec76}, left and middle panel), for respective total charges and running times of 62 (71) C and 232 (152) hours. 

%==============================================
%==============================================

The off-resonance yield was measured at $E_p$ = 188.0, 205.2, 250.0, and 310.0 keV, beam energies that were either below or well above the energies of known or supposed resonances, in order to study the contributions of  direct capture and by broad resonances (Figure~\ref{fig:Gammaspec76}, right panel). % db changed 06.06.2018

{\bf Results.} The three previously reported resonances \cite{Cavanna15-PRL,Depalo16-PRC,Cavanna15-PRL-Erratum,Kelly17-PRC,Lennarz17-NPA8} were confirmed with high statistics (Figure~\ref{fig:Gammaspectra}). 
For the determination of their resonance strengths $\omega\gamma$, the following formulation for the yield $Y(E_p)$ was used, assuming that the resonance has a Breit-Wigner shape $\sigma_{\rm BW}(E)$ and narrow total width $\Gamma$ \cite{Vodhanel84-PRC,Hale01-PRC}. 
The energy distribution $f_{\rm beam}(E,\sigma_{\rm stragg})$ of the beam is given by the accelerator energy spread \cite{Formicola03-NIMA} and energy straggling from SRIM \cite{Ziegler10-NIMB} ($\sigma_{\rm stragg}$). Finally, the known \cite{Ferraro18-EPJA} target density $n(\tilde{x})$ and detection efficiency $\eta(\tilde{x})$ (maximal $\eta(\tilde{x})$ values are 0.41-0.65 depending on the decay scheme) as a function of position $\tilde{x}$ (along the whole active target length $\Delta x$) are used:
\begin{eqnarray}
Y(E_p) & = & \int\limits_{\Delta x} d\tilde{x} \int\limits_{E=0}^{E=E_p} dE \nonumber \\ 
 & & \sigma_{\rm BW}(E) \; f_{\rm beam}(E,\sigma_{\rm stragg}(\tilde{x})) \; n(\tilde{x}) \; \eta(\tilde{x}) \label{eq:Yield}
\end{eqnarray}
The resultant yield from Eq.~(\ref{eq:Yield}) is 1-2\% lower than that obtained from the classical thick-target approximation \cite{Iliadis15-Book}, to be compared with 18-19\% effect in the LUNA-HPGe case \cite{Bemmerer18-EPL}. 
The final resonance strengths were determined in two independent analysis approaches: first, using the well-established LUNA adaptation of the GEANT3-based code \cite{Arpesella95-NIMA}, which includes all the above mentioned effects, and second, using GEANT4 \cite{Agostinelli03-NIMA} for the $\gamma$ detection efficiency and a subsequent analysis code in ROOT treating target profile, energy loss, and energy straggling. The two results \cite{Ferraro17-PhD,Takacs17-PhD} are found to be consistent and then averaged (Table~\ref{table:strengths}). 

%==============================================
%==============================================

The systematic uncertainty is dominated by the contributions from the $\gamma$-detection efficiency (5\%, including also the assumed branching ratio), the stopping power from SRIM (1.7\% \cite{Ziegler10-NIMB}), the beam intensity (1.5\%), and the effective target density (1.3\%, mainly given by the $\leq$5\% beam heating correction) \cite[for details]{Ferraro18-EPJA}. The statistical uncertainty is  below 2\% for the observed resonances and 1-9\% for the off-resonance data.

The suggested resonances at $E_p$ = 71 and 105\,keV, corresponding to $J^\pi$ = 1/2$^+$ $^{23}$Na levels at $E_{\rm x}$ = 8862 and 8894\,keV, respectively, were again not found here. Instead, new and much more stringent upper limits were determined for the 63-78 and 95-113\,keV energy ranges (Table~\ref{table:strengths}). This gain in sensitivity is illustrated in Figure~\ref{fig:Gammaspec76} (left and middle panels), where experimental spectra are shown together with a simulation assuming a strength equal to the previous upper limit \cite{Cavanna15-PRL,Depalo16-PRC,Cavanna15-PRL-Erratum,Bemmerer18-EPL}  and the decay pattern from the nearby $E_{\rm x}$ = 8830\,keV ($E_p$ = 37 keV) level which has the same spin-parity \cite{Firestone07-NDS23}.  % db changed 06.06.2018 and after Antonio 

%==============================================
\begin{figure}[tb]
%\centering
\includegraphics[width=\columnwidth]{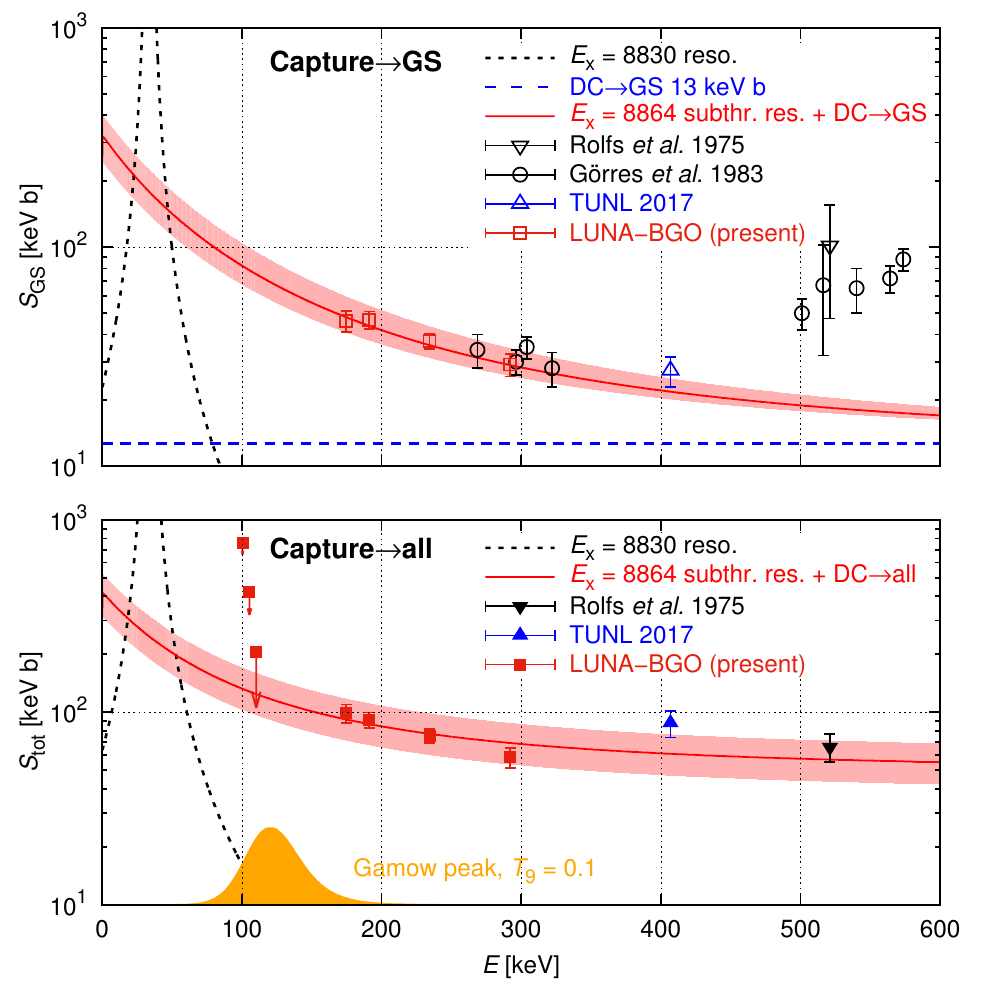}
\caption{Off-resonance capture in $^{22}$Ne($p,\gamma$)$^{23}$Na from the literature \cite{Rolfs75-NPA,Goerres83-NPA,Kelly17-PRC} and the present work. The result of the present fit is given by a red line with the shaded red area for the uncertainty. Top: Capture to the ground state in $^{23}$Na. Bottom: Total S-factor, and  Gamow peak for $T_9 = 0.1$.}
\label{fig:Offresonance}
\end{figure}
%==============================================

The off-resonance S-factor data (typical spectrum in Figure~\ref{fig:Gammaspec76}, right panel) rise towards lower energies. This is expected for a broad subthreshold resonance, such as the $J^\pi$=1/2$^+$ state at $E_{\rm x}$ = 8664\,keV ($E$ = -130\,keV), which decays (84$\pm$3)\% to the ground state \cite{Firestone07-NDS23}. Its spectroscopic factor and a constant ground state S-factor are fitted to match the ground state capture component of the present non-resonant data (Figure \ref{fig:Offresonance}, top panel). The fit gives $C^2S$=0.42$\pm$0.08, in between the previous values $C^2S$=0.30 \cite{Goerres83-NPA} and 0.58 \cite{Terakawa93-PRC}, and $S_{\rm GS}^{\rm DC}$ = 13$\pm$5\,keV\,b, consistent with previous work \cite{Goerres83-NPA}. The narrow low-energy resonance at $E_p$ = 37\,keV ($E_{\rm x}$ = 8830\,keV) contributes only negligibly 
\cite{Goerres83-NPA,Hale01-PRC}, and has been excluded from the fit (but plotted for reference in Figure \ref{fig:Offresonance}).  

In a second step, the total S-factor is fitted so that the sum of the total S-factor and the above determined subthreshold resonance match the present experimental total S-factor data  (Figure \ref{fig:Offresonance}, bottom panel). The result is $S_{\rm tot}(E)$ = (50$\pm$12)\,keV\,b, consistent with 62\,keV\,b \cite{Goerres83-NPA}, the previously accepted value \cite{NACRE99-NPA,Iliadis10-NPA841_251,Sallaska13-ApJSS}. 

For the thermonuclear reaction rate, the laboratory strengths shown in Table \ref{table:strengths}  have to be divided by the electron screening enhancement factor, calculated for the adiabatic limit \cite{Assenbaum87-ZPA}. 

%%%%%%%%%%%%%%%%%%%%%%%%%%%%%%%%%%%%%%%%%%%%%%%%%%%%%
\begin{table*}[tb]
\caption{Strengths of low-energy resonances in the $^{22}$Ne($p,\gamma$)$^{23}$Na reaction. Upper limits are given for 90\% confidence level.}
\begin{center}
\label{table:strengths}
\begin{tabular}{ | D{.}{.}{1} c | r@{$\times$}l | r | c | r@{$\times$}l |r|}
\hline
\multicolumn{2}{|c}{\textbf{Energy\,[keV]}} & \multicolumn{6}{c|}{\textbf{Strength} $\omega\gamma$\,[eV]} & Screening \\
\multicolumn{1}{|c|}{$E_{\rm p}^{\rm res}$} & $E_{\rm x}$ & \multicolumn{2}{c|}{Iliadis {\it et al.}\,\cite{Iliadis10-NPA841_251}} & LUNA-HPGe \cite{Cavanna15-PRL,Depalo16-PRC,Cavanna15-PRL-Erratum,Bemmerer18-EPL} & TUNL \cite{Kelly17-PRC} & \multicolumn{2}{c|}{LUNA-BGO (present)} & enhancement \\
\hline
\hline
37    & 8830 & [3.1\,$\pm$\,1.2] & 10$^{-15}$ & - & - & \multicolumn{2}{c|}{-} & - \\ 
71    & 8862 & \multicolumn{2}{c|}{-} & $\leq$\,1.5$\times$10$^{-9}$ & - & $\leq$\,6 & 10$^{-11}$ & 1.266 \\
105   & 8894 & \multicolumn{2}{c|}{-} & $\leq$\,7.6$\times$10$^{-9}$ & - & $\leq$\,7 & 10$^{-11}$ & 1.140 \\
156.2 & 8944 & [9.2$\pm$3.0] & 10$^{-9}$& [1.8\,$\pm$\,0.2]$\times$10$^{-7}$ & [2.0\,$\pm$\,0.4]$\times$10$^{-7}$ & [2.2\,$\pm$\,0.2] & 10$^{-7}$ & 1.074 \\

189.5 & 8975 & $\leq$2.6 & 10$^{-6}$ & [2.2\,$\pm$\,0.2]$\times$10$^{-6}$ & [2.3\,$\pm$\,0.3]$\times$10$^{-6}$ & [2.7\,$\pm$\,0.2] & 10$^{-6}$ & 1.055\\

215   & 9000 &  \multicolumn{2}{c|}{-} & $\leq$\,2.8$\times$10$^{-8}$ & - & \multicolumn{2}{c|}{-} & 1.045 \\
259.7 & 9042 & $\leq$\,1.3 & 10$^{-7}$ & [8.2\,$\pm$\,0.7]$\times$10$^{-6}$ & - & [9.7\,$\pm$\,0.7] & 10$^{-6}$ & 1.034 \\

\hline          
\end{tabular}
\end{center}
\end{table*}%
%%%%%%%%%%%%%%%%%%%%%%%%%%%%%%%%%%%%%%%%%%%%%%%%%%%%

{\bf Discussion.} The present strengths for the three recently reported resonances at $E_p$ = 156.2, 189.5, and 259.7 keV are higher than those from the LUNA-HPGe experiment \cite{Cavanna15-PRL,Depalo16-PRC,Cavanna15-PRL-Erratum,Bemmerer18-EPL} but consistent within 2$\sigma$ (Table~\ref{table:strengths}). For the LUNA-HPGe data, the yield was only measured at two angles, and weak branches may have been missed \cite{Cavanna15-PRL,Depalo16-PRC,Cavanna15-PRL-Erratum,Bemmerer18-EPL}. The present near 4$\pi$ geometry limits effects even from strongly anisotropic angular distributions to $\leq$4\%, and also weak branches are included in the add-back spectrum. The present results supersede those of Refs.~\cite{Cavanna15-PRL,Depalo16-PRC,Cavanna15-PRL-Erratum,Bemmerer18-EPL}. The present strengths are higher than TUNL \cite{Kelly17-PRC} but consistent within 1$\sigma$. 

\begin{figure}[tb]
%\centering
\includegraphics[width=\columnwidth]{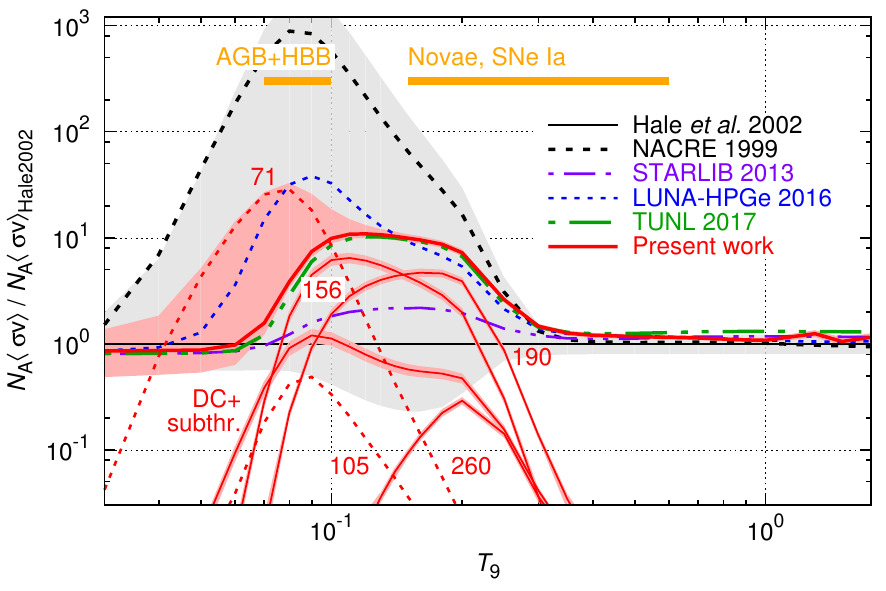}
\caption{Thermonuclear reaction rate for the $^{22}$Ne($p,\gamma$)$^{23}$Na reaction from previous evaluations \cite{NACRE99-NPA,Sallaska13-ApJSS} and from experiments: LUNA-HPGe \cite{Cavanna15-PRL,Depalo16-PRC,Cavanna15-PRL-Erratum,Bemmerer18-EPL}, TUNL \cite{Kelly17-PRC}, present work. The rates are shown relative to the frequently used \cite{DellAgli18-MNRAS,Ventura18-MNRAS,Karakas18-MNRAS} rate by Hale {\it et al.} \cite{Hale01-PRC}. 
The individual contributions from the present work are labeled. 
For the total rates by Hale {\it et al.} and from the present work, as well as the individual contributions from the present work, shaded error bands have been added. 
The contribution from the 37 keV resonance dominates at low temperature and has been omitted for clarity.
}
\label{fig:Rate}
\end{figure}

In the present thermonuclear reaction rate (Figure~\ref{fig:Rate}), for the first time all relevant processes at low energy, in particular non-resonant capture and stringent upper limits on the two suggested resonances at $E_p$ = 71 and 105 keV, are based on experiment. The error is higher than that of the TUNL rate \cite{Kelly17-PRC}, because TUNL does not take any uncertainty into account due to the two suggested resonances, whereas in the present work, experimental upper limits have been used. 

The new upper limits for the $E_p$ = 71 and 105\,keV resonances are 25 and 110 times lower, respectively, than those from the LUNA-HPGe experiment.
As a consequence, these resonances, which produced a bump in the NACRE \cite{NACRE99-NPA} reaction rate and caused some remaining ambiguity in the LUNA-HPGe data \cite{Cavanna15-PRL,Depalo16-PRC,Cavanna15-PRL-Erratum,Bemmerer18-EPL} now play almost no role  from the astrophysical point of view (Figure~\ref{fig:Rate}). 
A much higher contribution is observed for the non-resonant component, which makes up to 30\% (at $T_9$ = 0.08) of the new, total rate. At even lower temperatures, $T_9$ $<$ 0.07, the $E_p$ = 37\,keV resonance dominates. Its indirectly derived strength is 10$^{-15}$\,eV \cite{Hale01-PRC}, too low for a direct measurement regardless of background. 

In order to gauge the impact of the new rate for AGB stars that experience hot-bottom burning, two characteristic stellar models have been run: First, the model in Ref.~\cite{Karakas18-MNRAS} has been re-calculated twice using, first, the rate by Hale {\it et al.} \cite{Hale01-PRC}, and second, the rate from the present work, which is ten times larger than Hale at a typical HBB temperature of $T_9$ = 0.1. For the case of a $M$ = 6$M_\odot$ star with metallicity\footnote{The shorthand [X/Y] with X, Y chemical elements, or groups thereof, refers to the decadic logarithm of a double ratio: The ratio of abundances of elements X and Y in the star under study, divided by the same ratio in the Sun.} [Fe/H] = -0.7, the surface sodium abundance at the end of ~40 thermal pulses increased by a factor of 1.3, from [Na/Fe] = 0.07 to [Na/Fe] = 0.18. 
The abundances of oxygen and all other nuclei studied did not vary significantly, except for a slight decrease in [Ne/Fe]. The increased [Na/Fe] brings the model closer to the pattern observed in 47 Tucanae \cite{Carretta09-AA}. -- Second, a set of models with initial masses $M$ = 3-5$M_\odot$, initial metallicity [Fe/H] = -1.4,
%$Z$ = 0.0005, 
and an $\alpha$-element-enhanced mixture with [$\alpha$/Fe]=0.4 from Ref.~\cite{Slemer17-MNRAS} was re-calculated using the COLIBRI code \cite{Marigo13-MNRAS}, again first with the Hale {\it et al.} rate \cite{Hale01-PRC}, and then the present rate. For stars with initial masses $\geq4M_\odot$, the new rate again leads to much higher ejected $^{23}$Na mass, e.g. in the 5$M_\odot$ case a factor of two more $^{23}$Na, affected by a factor of three lower uncertainty when compared to the old rate.

{\bf Conclusion.} Based on the present new, high-luminosity and low-background data, the remaining uncertainties on the $^{22}$Ne($p,\gamma$)$^{23}$Na reaction rate at low temperature have been greatly reduced. In particular, the discrepancy between the  rates by NACRE \cite{NACRE99-NPA} and by Hale/Iliadis/STARLIB \cite{Hale01-PRC,Iliadis10-NPA841_31,Sallaska13-ApJSS}  is overcome by the present results. In the relevant temperature range, compared with the present rate, the NACRE \cite{NACRE99-NPA} rate is high by a factor of 10-100, and the frequently used Hale {\it et al.} \cite{Hale01-PRC} rate is low by a factor of 10. 

The new rate has been shown to enhance $^{23}$Na production in the HBB process. Thus it may help to explain the Na-O anticorrelation in globular cluster stars \cite{Carretta09-AA}. There may be further consequences of the new rate on sodium production in carbon-enhanced metal poor stars, depending on the scenario of hydrogen burning \cite{Choplin16-AA}. Overall, the new precise nuclear physics input will be instrumental in future studies of stellar scenarios \cite{Slemer17-MNRAS,Buntain17-MNRAS,Choplin17-AA,DellAgli18-MNRAS,Ventura18-MNRAS,Karakas18-MNRAS} addressing hot-bottom burning. 

%==============================================
%==============================================

% ===========================================================
\begin{acknowledgments}
Financial support by INFN, DFG (BE 4100-4/1), the Helmholtz Association (NAVI VH-VI-417 and ERC-RA-0016), NKFIH (K120666), MTA Lend\"ulet (2014-17), the COST Association (ChETEC, CA16117), and DAAD fellowships at HZDR for F.C. and R.D. are gratefully acknowledged. 
\end{acknowledgments}
% ===========================================================
% BibTeX users please use
% \bibliography{Danielsbib}

\end{document}